\documentclass[twocolumn]{aastex62}

\providecommand\rh{\mbox{$r_{\mathrm{h}}$}}
\providecommand\afrho[1][\theta]{\mbox{$A(#1)f\rho$}}
\providecommand\afr{\mbox{$Af\rho$}}

\usepackage{amsmath}

\usepackage{CJK}

\graphicspath{{./}{figures/}}

\received{July 16, 2019}
\revised{October 22, 2019}
\accepted{November 4, 2019}
\submitjournal{ApJ Letters}
\shortauthors{Kelley et al.}

\begin{document}
\begin{CJK*}{UTF8}{gbsn}

\title{Comet 240P/NEAT is Stirring}

\author[0000-0002-6702-7676]{Michael S. P. Kelley}
\affil{Department of Astronomy, University of Maryland, College Park, MD 20742-0001, U.S.A.}
\email{msk@astro.umd.edu}

\author[0000-0002-2668-7248]{Dennis Bodewits}
\affiliation{Physics Department, Leach Science Center, Auburn University, Auburn, AL 36832, U.S.A.}

\author[0000-0002-4838-7676]{Quanzhi Ye (叶泉志)}
\affiliation{Division of Physics, Mathematics, and Astronomy, California Institute of Technology, Pasadena, CA 91125, U.S.A.}
\affiliation{Infrared Processing and Analysis Center, California Institute of Technology, Pasadena, CA 91125, U.S.A.}

\author[0000-0002-4767-9861]{Tony L. Farnham}
\affil{Department of Astronomy, University of Maryland, College Park, MD 20742-0001, U.S.A.}

\author[0000-0001-8018-5348]{Eric C. Bellm}
\affiliation{DIRAC Institute, Department of Astronomy, University of Washington, 3910 15th Avenue NE, Seattle, WA 98195, U.S.A.}

\author{Richard Dekany}
\affiliation{Caltech Optical Observatories, California Institute of Technology, Pasadena, CA 91125, U.S.A.}

\author[0000-0001-5060-8733]{Dmitry A. Duev}
\affiliation{Division of Physics, Mathematics, and Astronomy, California Institute of Technology, Pasadena, CA 91125, U.S.A.}

\author[0000-0003-3367-3415]{George Helou}
\affiliation{Infrared Processing and Analysis Center, California Institute of Technology, Pasadena, CA 91125, U.S.A.}

\author[0000-0002-6540-1484]{Thomas Kupfer}
\affiliation{Kavli Institute for Theoretical Physics, University of California, Santa Barbara, CA 93106, USA}

\author[0000-0003-2451-5482]{Russ R. Laher}
\affiliation{Infrared Processing and Analysis Center, California Institute of Technology, Pasadena, CA 91125, U.S.A.}

\author[0000-0002-8532-9395]{Frank J. Masci}
\affiliation{Infrared Processing and Analysis Center, California Institute of Technology, Pasadena, CA 91125, U.S.A.}

\author[0000-0002-8850-3627]{Thomas A. Prince}
\affiliation{Division of Physics, Mathematics, and Astronomy, California Institute of Technology, Pasadena, CA 91125, U.S.A.}

\author[0000-0001-7648-4142]{Ben Rusholme}
\affiliation{Infrared Processing and Analysis Center, California Institute of Technology, Pasadena, CA 91125, U.S.A.}

\author[0000-0003-4401-0430]{David L. Shupe}
\affiliation{Infrared Processing and Analysis Center, California Institute of Technology, Pasadena, CA 91125, U.S.A.}

\author[0000-0001-6753-1488]{Maayane T. Soumagnac}
\affiliation{Benoziyo Center for Astrophysics, Weizmann Institute of Science, Rehovot, Israel}

\author{Jeffry Zolkower}
\affiliation{Caltech Optical Observatories, California Institute of Technology, Pasadena, CA 91125, U.S.A.}

\begin{abstract}
  Comets are primitive objects that formed in the protoplanetary disk, and have been largely preserved over the history of the Solar System.  However, they are not pristine, and surfaces of cometary nuclei do evolve.  In order to understand the extent of their primitive nature, we must define the mechanisms that affect their surfaces and comae.  We examine the lightcurve of comet 240P/NEAT over three consecutive orbits, and investigate three events of significant brightening ($\Delta m\sim-2$~mag).  Unlike typical cometary outbursts, each of the three events are long-lived, with enhanced activity for at least 3 to 6 months.  The third event, observed by the Zwicky Transient Facility, occurred in at least two stages.  The anomalous behavior appears to have started after the comet was perturbed by Jupiter in 2007, reducing its perihelion distance from 2.53 to 2.12~au.  We suggest that the brightening events are temporary transitions to a higher baseline activity level, brought on by the increased insolation, which has warmed previously insulated sub-surface layers.  The new activity is isolated to one or two locations on the nucleus, indicating that the surface or immediate sub-surface is heterogeneous.  Further study of this phenomenon may provide insight into cometary outbursts, the structure of the near-surface nucleus, and cometary nucleus mantling.
\end{abstract}


\section{Introduction} \label{sec:intro}

Cometary nucleus surfaces are dynamic, with many processes affecting their volatile content, strength, particle size distribution, and mass-loss \citep{veverka13, thomas13-tempel1, thomas15-morph, elmaarry15, sunshine16}.  Most processes are ultimately driven by insolation.  This fact enables the study of nuclear surfaces through examination of gas and dust production as they rotate and orbit the Sun.  The correlation of composition or mass-loss rates with insolation may reveal the composition, structure, or evolution of the near-surface layer \citep[e.g.,][]{biver97, meech13-ison, feaga14, bodewits14-garradd}.

Gradual and repeated variations occur on seasonal and diurnal timescales as localized sources on the nucleus vary in activity \citep[e.g.,][]{hayward00, schleicher07, ahearn11, kramer17-cg}.  In contrast, cometary outbursts are more stochastic.  These impulsive increases in mass-loss rate eject material into the coma, causing an immediate and rapid brightening in telescopic observations.  The total brightness of the coma varies with a near-exponential decay as the outburst ejecta slowly leaves the vicinity of the nucleus \citep{hughes90}.  The causes of outbursts vary \citep{hughes91}.  For example, they may be driven by sub-surface energy storage \citep{agarwal17-outburst}, rotationally induced mass shedding \citep{steckloff18}, cliff collapse \citep{pajola17}, and water ice phase state transitions \citep{prialnik90, belton09}.  On occasion, outbursts signal the complete disruption of the nucleus \citep{farnham01, knight14-ison-perihelion, li15-c2010x1}.

Comet \object{240P/NEAT} is a Jupiter-family comet, discovered in 2002 as P/2002~X2 by the Near-Earth Asteroid Tracking (NEAT) survey with the 1.2-m Samuel Oschin telescope at Palomar Observatory \citep{lawrence02-iauc8029}.  With a 7.6-yr orbital period, it has been observed over 3 perihelion passages.  On 2007 July 10, it made a close approach to Jupiter ($\Delta_J=0.25$~au, heliocentric distance \rh=5.5~au) and its perihelion distance, $q$, dropped from 2.53 to 2.12~au (NASA JPL Small-Body Database, orbital solution JPL K182/8), corresponding to a 40\% increase in insolation at perihelion.  Prior to this encounter, the comet's orbit had been stable with $q$ near 2.5--2.6~au for at least 80 years, according to the same JPL solution.

After passing through perihelion in the newly perturbed orbit, an apparent 2-mag outburst was reported by B.\ Haeusler\footnote{\url{https://groups.yahoo.com/neo/groups/comets-ml/conversations/messages/17241}}, occurring between 2011 March 29 and April 06.  On the next perihelion passage, a second apparent 2-mag outburst was reported by \citet{sato17-cbet4427}, between 2017 July 18.63 and August 28.59 UTC.  In 2018, S.\ Yoshida (personal communication) received a report from T.\ Ikemura and H.\ Sato (at Shinshiro, IAU observatory code Q11) that the comet had experienced a third apparent outburst, 1--2~mag in strength, between 2018 November 14.81 and December 12.68 UTC.

We examine the apparent outbursts and baseline activity of comet 240P.  We present photometry of the comet from the Zwicky Transient Facility \citep[ZTF;][]{bellm19-ztf, graham19-ztf}, the Palomar Transient Factory \citep[PTF;][]{rau09-ptf, law09-ptf}, and the NEAT survey.  These data, in combination with photometry reported to the Minor Planet Center (MPC), reveal a comet in repeated transition between two different activity states.


\section{Observations and Results}\label{sec:obs}

The ZTF is a time-domain all-sky survey and successor to the PTF. First light was acquired 2017 November 1, and science operations commenced 2018 March 20.  The camera utilizes 16 6k$\times$6k CCDs (1.01\arcsec{} per pixel) to cover a 47 square degree field of view.  It is mounted on the 1.2-m Oschin Schmidt telescope at Palomar Observatory.  Survey operations typically use 30-s exposures, allowing ZTF to cover 3800 square degrees an hour to a 5$\sigma$ depth of $r=20-21$~mag \citep{graham19-ztf, bellm19-ztf}.

We searched for observations of comet 240P/NEAT in the ZTF Data Release 1 and Partnership data archives \citep{masci19-ztf} with the \texttt{ZChecker} program \citep{Kelley2019}.  Survey coverage of the comet began on 2018 September 11.51 UTC at $\rh=2.3$~au, 119 days after perihelion ($T_P=$ 2018 May 15.88 UTC).  We inspected 63 ZTF $g$-, $r$-, and $i$-band images covering the comet, and measured its brightness using 15,000-km radius apertures.  Several images were dropped from the analysis for various reasons, including stellar contamination, background artifacts, suspected clouds, or high background.  The uniform aperture size was chosen to account for seeing and geocentric distance variations throughout the observation period (minimum aperture is 6.9\arcsec, median seeing is 2.2\arcsec).  Photometry was calibrated to the PanSTARRS (PS1) DR1 catalog \citep{tonry12-ps1} using the ZTF pipeline \citep{masci19-ztf}.  We measured $g-r=0.56\pm0.02$~mag from the average of seven $g$- and $r$-band image pairs; $r-i$ cannot be directly measured because the $i$-band images are separated from the other images by many days.  We assume a constant spectral gradient from $g$ to $i$, i.e., $r-i=0.27$~mag.  We used these values to color correct the data from the ZTF filters to the PS1 system (AB magnitudes).  The results are binned by day (Table~\ref{tab:phot}).

We also searched for comet 240P in the PTF archive with an online application at the Infrared Science Archive, and found 8 images observed with an $R$-band filter.  PTF image processing and photometric calibration is described by \citet{laher14-ptf} and \citet{ofek12-ptf}.  We calibrated the images to PS1 $r$-band magnitudes using background stars and the \texttt{calviacat} program \citep{kelley19-calviacat}.  Photometry of the comet in 9.5\arcsec{} radius apertures is presented in Table~\ref{tab:phot}.  The fixed angular size was chosen to make the results more comparable to the MPC photometry (justified in Section~\ref{sec:2018}).

In addition, we obtained all comet 240P photometry reported to the MPC \citep{Williams2019}.  The data were taken with a wide range of calibration methods, aperture sizes, and bandpasses.  As a result, there is a large scatter in reported magnitudes, even when data are separated into ``nuclear'' (small) and ``total'' (whole coma) magnitudes.  We select all photometry from a subset of observatories (360, 644, 693, 699, 704, E12, H45, G96, T08, T05, V06, 958, B96, H47, J38, B82, and A71), chosen for broad time coverage and the best-quality data.  The remaining MPC data produce an improved lightcurve, but still have scatter at the magnitude level.  However, activity trends are apparent in the data, therefore we include them in our analysis.

We also searched the data archived at the Canadian Astronomy Data Centre \citep{Gwyn2012} for pre-discovery (2002) detections of 240P. The comet was covered by NEAT survey images on 1998 May 03 and May 24 ($\rh=5.4$~au), but the predicted brightness ($V\sim21$~mag) was below the sensitivity limit of the images ($V\sim19$~mag). A search by eye on the images within the uncertainty ellipse ($<3''$) did not turn up any evidence for the object. We conclude that the comet was not more than 2 magnitudes brighter than the 2003 activity level at that time.

Three images of comet 240P on 2003 January 16 were found in the partial NEAT data archive of \citet{bauer13-neat}.  We bias-subtracted and flat fielded the data, and measured the coma within 7\arcsec{} radius apertures.  The photometric aperture is limited due to a nearby star.  The unfiltered images were calibrated to PS1 $r$-band magnitudes using background stars.  The weighted-mean photometry is presented in Table~\ref{tab:phot}.


\section{Models}
To model the comet's photometric behavior, we use the \afr{} coma quantity of \citet{ahearn84-bowell}.  It is proportional to the apparent brightness of the comet, and is defined as the product of grain albedo, filling factor within the aperture, and aperture size projected to the distance of the comet:
\begin{equation}
  \afrho = \frac{4 \Delta^2 \rh^2 F_\lambda}{\rho S_\lambda},
  \label{eq:afrho}
\end{equation}
where $A(\theta)f\rho$ specifies that the measurement is for a specific phase angle $\theta$, $\Delta$ is the observer-comet distance, $F_\lambda$ is the observed spectral flux density of the coma within a circular aperture with projected linear radius $\rho$, and $S_\lambda$ is the spectral flux density of sunlight at 1 au.  Despite the units of length, \afr{} is a proxy for the comet's intrinsic dust coma activity, i.e., mass loss rate \citep{fink12}.  To model the comet's brightness, we assume \afr{} varies as a power-law based on the heliocentric distance (\rh):
\begin{equation}
  \afrho = \afrho[0\degr]\ \Phi(\theta) \left(\frac{\rh}{q}\right)^k,
\end{equation}
where $\Phi$ is a phase function for light scattered by cometary dust \citep{schleicher11}, $q$ is the perihelion distance, and $k$ is the power-law slope.  \afrho[0\degr] is the value that would be measured if the comet were observed at a phase angle of 0\degr.

We also interpret the comet's activity state with the ice sublimation model of \citet{cowan79}.  This model balances absorption of sunlight by a low-albedo (5\% bond albedo) spherical nucleus with the energy losses of thermal radiation and ice sublimation.  Based on spacecraft observations, cometary surfaces have low thermal inertias \citep{groussin13, davidsson13-thermali, gulkis15}, i.e., their surface temperatures are in near-instantaneous equilibrium with sunlight.  We assume the same property for the nucleus of 240P in our ice sublimation model.

\begin{figure*}
    \plotone{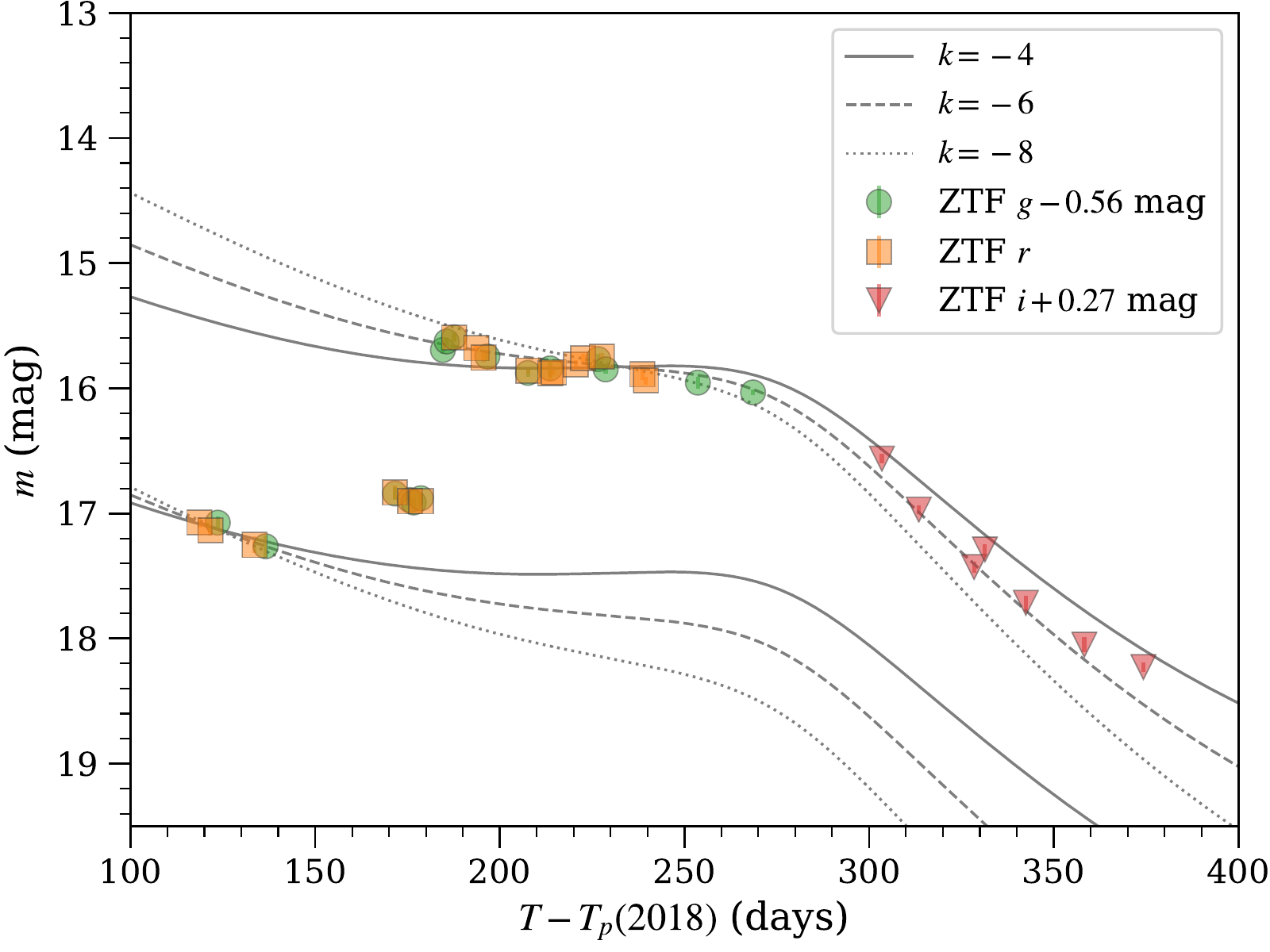}
    \caption{ZTF photometry of comet 240P/NEAT, measured with a 15,000-km radius aperture.  Several photometric models are shown to demonstrate our model uncertainty on the heliocentric distance power-law slope, $k$.}
    \label{fig:ztf}
\end{figure*}


\section{Analysis}
\subsection{2018 Orbit: $T_P$=2018 May 15.88}\label{sec:2018}
The ZTF photometry yields an unusual lightcurve with two apparent brightening events at $T_P+172$ and +185 days (Fig.~\ref{fig:ztf}).  The first brightening had a strength of $\Delta m\sim-0.7$~mag, and occurred between $T_P+136.6$ and $+171.6$~days (September 29 to November 03), where $T_P$ is the perihelion date.  After seven nights of a near-constant apparent magnitude, a second event occurred between $T_P+178.7$ and $+184.6$~days (November 10 and 16), increasing the total brightening to $\Delta m=-2$~mag.  The latter event is the 2018 outburst identified by Ikemura and Sato (between $T_P+182.9$ and +210.8~days).  These data indicate that the full 2-mag event was not simply a gradual increase in activity, but occurred in at least two stages.  The apparent magnitude peaks on day three of the second event at $+187.6$~days.  Subsequent structure in the lightcurve suggests a $\Delta m\sim-0.1$~mag peak near $T_P+221$~days.

Figure~\ref{fig:ztf} presents lightcurves based on our \afr{} model using the heliocentric distance slopes $k=-4$, $-6$, and $-8$.  The lightcurves have the same aperture radius as the ZTF photometry (15,000~km).  The data at $\sim T_P+130$~days cannot be used to discriminate between the three slopes (reduced $\chi^2$ values are 0.4--0.6).  However, $k=-6$ is the best of the three, and predicts $\afrho[0\degr]=213$~cm at perihelion.

None of our models account for the $\sim2$~mag increase in activity at $\sim T_P+180$~days.  Comparing this change in brightness to the $k=-4$ lightcurve implies the comet had a very slow return to quiescence, but even at $T_P+350$~days the comet is still $-1.3$~mag brighter than the model.  The $k=-8$ lightcurve suggests the post-event brightening increased with time, up to 2.8~mag by $T_P+350$~days.  Although there is no \textit{a priori} requirement that any of these models fit the post-event lightcurve, the middle value, $k=-6$, is most consistent with it, and we adopt this slope for the remainder of the paper.

All three model lightcurves are nearly parallel to the ZTF photometry, which suggests that the event around $T_P+180$~days was not a typical outburst, but rather a new sustained activity level.  Good agreement with the data is obtained with the $k=-6$ model for $\afrho[0\degr](q)=1346$~cm (Fig.~\ref{fig:ztf}), equivalent to a factor of 6 increase in the dust production rate.

\begin{figure*}
  \plotone{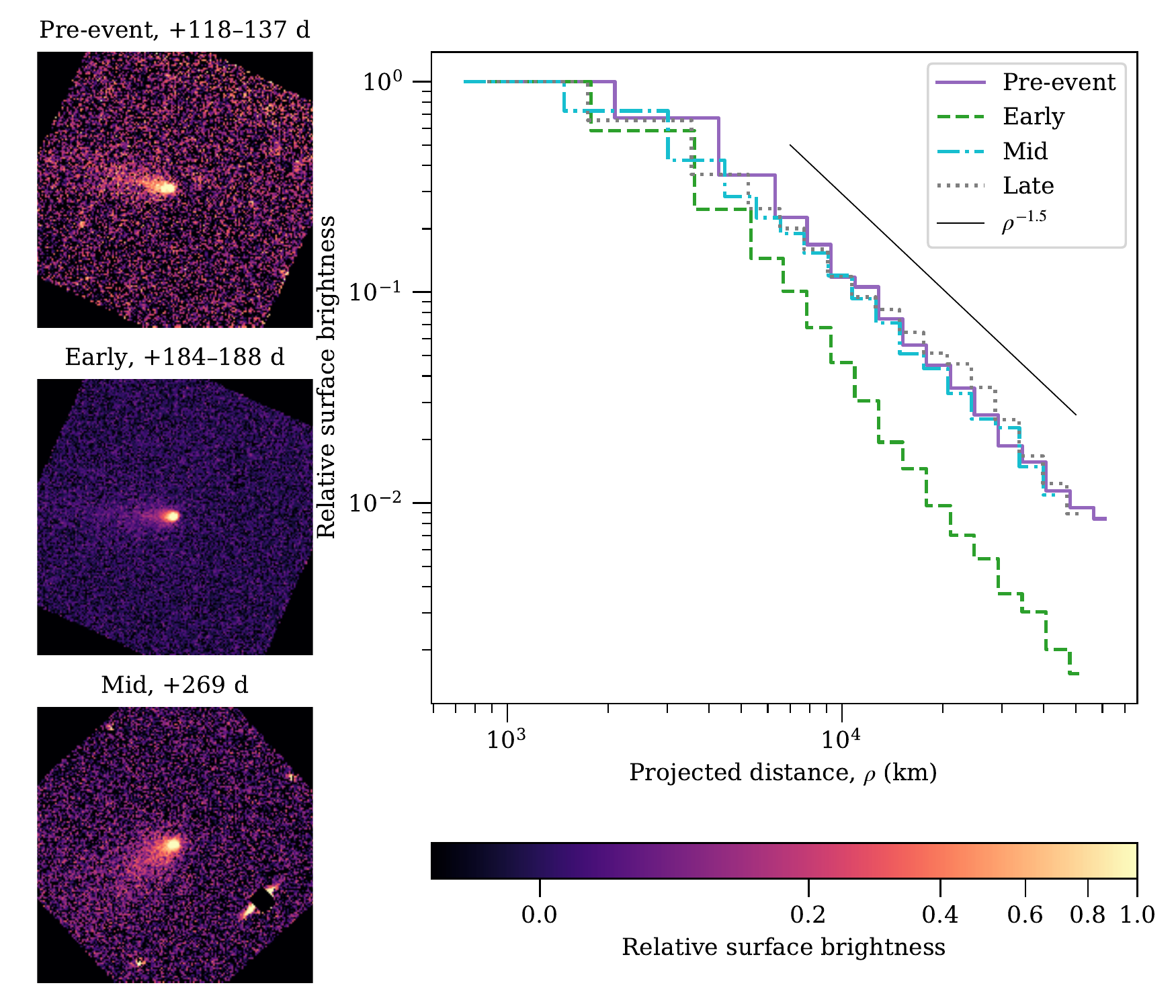}
  \caption{(Right) ZTF images of comet 240P from the 2018 event (times are with respect to perihelion; see text for details).  The images are displayed normalized to the surface brightness in an 8\arcsec{} radius aperture and the projected velocity vector is to the right.  (Left) Radial profiles for each image, plus an additional image from the late-event data ($T-T_p=303-328$~days).  Note the first three bins are 1, 2, and 3\arcsec{}.}
  \label{fig:images}
\end{figure*}

We examined the ZTF images to determine if the sustained brightening was due to lingering large grains, a new fragment, or to new activity.  Images were averaged into three bins to enhance the data quality (pre-event: 2018~September~11 to 29 (5 images); early-event: November~16 to 19 (5); mid-event: 2019~February~02 (1); late-event: March~15 to April~09 (3)).  The results and azimuthally averaged profiles are presented in Fig.~\ref{fig:images}.  There is no morphological evidence for a new fragment.  Before the event, the radial profile was close to $\rho^{-1.5}$, the canonical distribution of a tail-dominated image \citep{jewitt87-profiles}.  Immediately after the event, the profile was steeper than $\rho^{-1.5}$, indicative of an outburst early in its evolution when the ejecta is close to the nucleus.  The radial profile returns to the pre-event distribution in the mid- and late-event images.  An impulsive event cannot simultaneously have a $\rho^{-1.5}$ profile and a consistently high intrinsic brightness over this 130 day period, unless it were accompanied with new activity.

\begin{figure*}
  \plotone{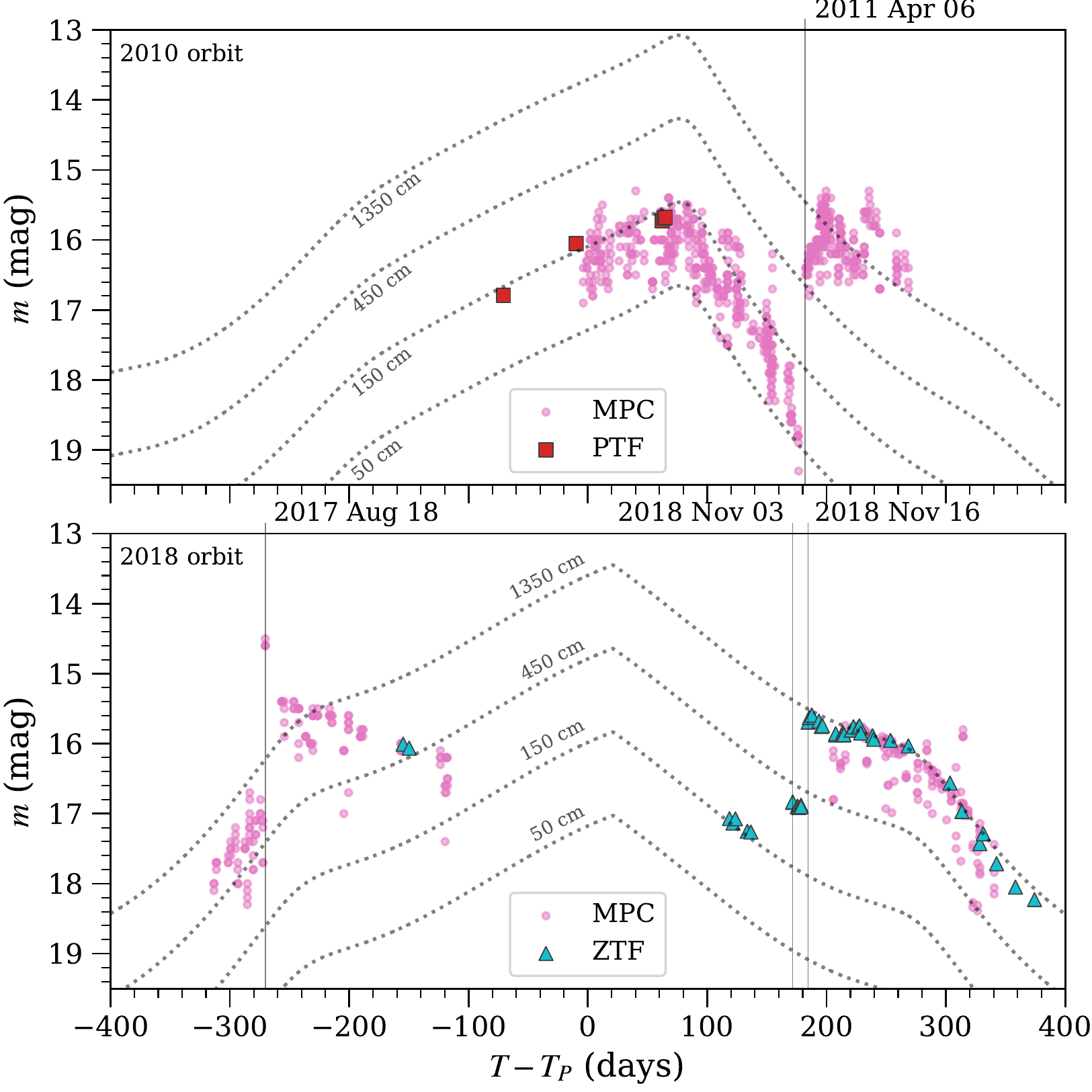}
  \caption{Lightcurve of comet 240P/NEAT versus perihelion time ($T-T_P$) in 2010--2011 (top) and 2017--2018 (bottom).  ZTF and PTF data are measured with 9.5\arcsec{} radius apertures.  Select photometry contributed to the MPC are also shown (see \ref{sec:obs} for details).  Model lightcurves are presented (dotted lines), based on the \afr{} formalism using the indicated value at perihelion (9.5\arcsec{} aperture, \rh$^{-6}$ scale factor).  Approximate epochs of transition to increased activity levels are marked with vertical lines.
  \label{fig:phot}}
\end{figure*}

Figure~\ref{fig:phot} presents the long-term lightcurve of the comet based on the ZTF, PTF, and MPC photometry.  Here, the ZTF photometry has been remeasured with a 9.5\arcsec{} radius aperture.  This choice of aperture size produced photometry in agreement with the data reported by the Asteroid Terrestrial-impact Last Alert System (ATLAS) survey  \citep{tonry18-atlas}, which accounts for most of the 2017/2018 photometry and informed our PTF photometric aperture choice.  For this aperture size, the 2018 events increased the coma \afr{} by a factor of $\sim9$.  The discrepancy with Fig.~\ref{fig:ztf}, where a factor of $\sim6$ change was found, is due to the limitations of the \afr{} model, which assumes a $1/\rho$ surface brightness distribution.

Inspection of the pre-perihelion MPC data reveals the apparent outburst identified by \citet{sato17-cbet4427} began between two sets of observations by ATLAS, at $T_P-272.3$ and $-270.3$~days (2018 August 16 and 18). The ATLAS photometry suggest an outburst strength of 2--3~mag, followed by a 1 to 1.5-mag decay in 14~days.  However, no ATLAS photometry exists during this decay period. After the initially rapid fading, the coma takes around 80~days to reach the pre-outburst activity level, and continues to fade through the last data at 150~days after the outburst.  This timescale is an order of magnitude longer than a typical cometary outburst \citep{hughes90}.  If the first ATLAS photometry points can be confirmed, it appears the event started with a typical outburst, but ended with an usually long fading period.

The pre- and post-perihelion events are separated by 220\degr{} of true anomaly.  Based on solar illumination of the nucleus, it is possible that a single active area is responsible for both.


\subsection{2010 Orbit: $T_P$=2010 October 04.27}
An activity model with \afrho[0\degr]=150~cm at perihelion has good agreement with our PTF photometry, and parallels most of the MPC data in 2010--2011 (Fig.~\ref{fig:phot}). The outburst reported by Haeusler is clear in the MPC lightcurve at $T_P+182$~days.  The data suggest the comet took 20~days to reach peak brightness, after which the coma remains near the 1350-cm model for 90~days, up to the last reported data for this period.  This is the same part of the orbit as the 2018 brightening; both have on-set dates near $+180$~days.  The portion of the orbit that covers the pre-perihelion event ($T_P-270$ to $-120$~days) was not observed.


\subsection{2003 Orbit: $T_P$=2003 March 29.61}
\begin{figure*}
  \plotone{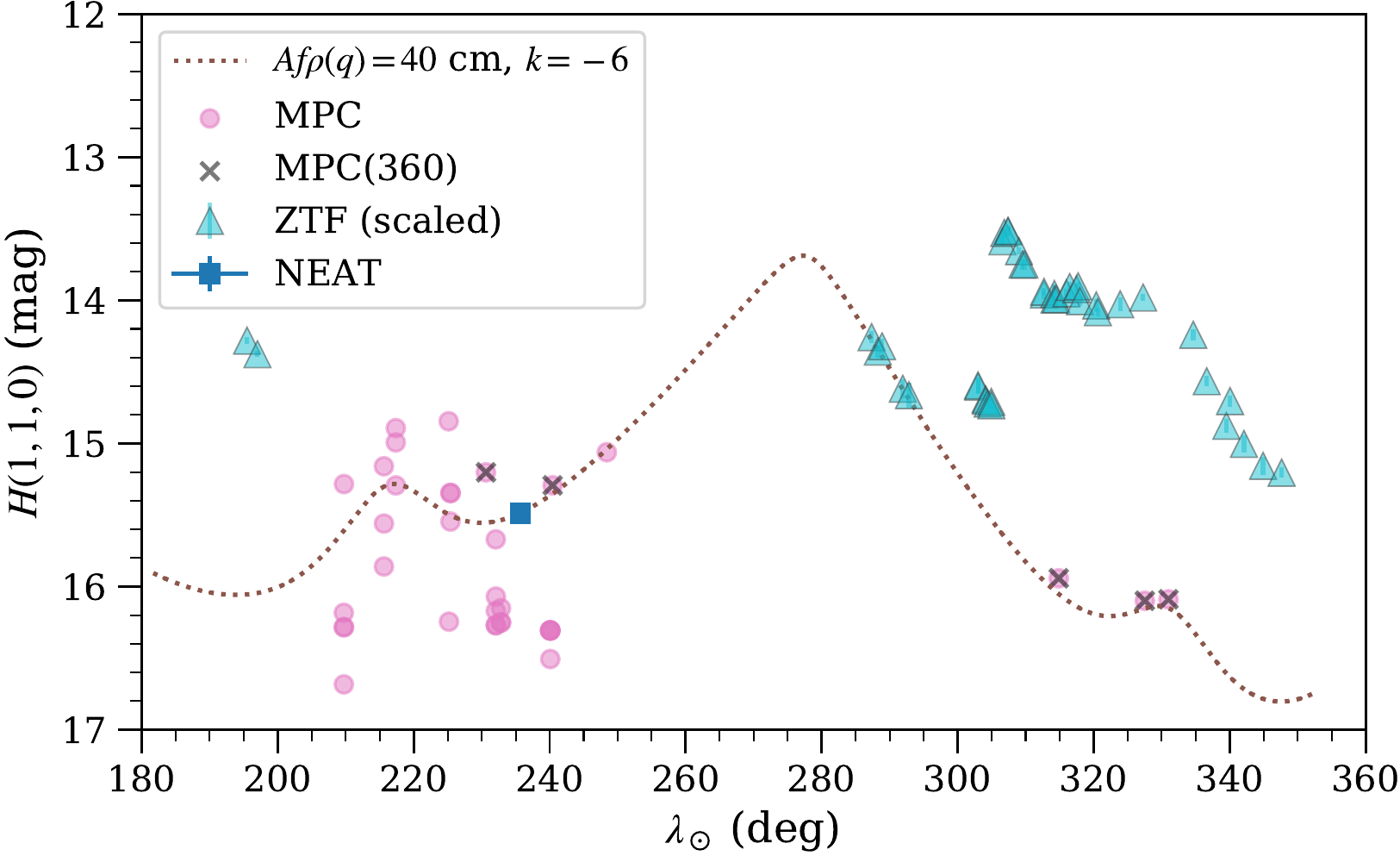}
  \caption{Absolute magnitude of comet 240P/NEAT from 2003 versus ecliptic longitude of the comet-Sun vector.  NEAT photometry within a 7\arcsec{} radius aperture, and select MPC photometry is shown.  Photometry from Kuma Kogen Astronomical Observatory (MPC code 360), which observed both sides of perihelion, is highlighted.  ZTF photometry from 2017--2018 is also shown, scaled to match the observation circumstances of 2003.  The model lightcurve (7\arcsec{} aperture, \rh$^{-6}$ scale factor) was fit to the NEAT data point.}
  \label{fig:2003}
\end{figure*}

The MPC photometric coverage in 2002--2004 covers $T_P-180$ to $-60$~days, and $+247$ to $+353$~days.  We examined these data for evidence of the 2011, 2017, and 2018 events.  Because the orbit changed after 2003, we plot the lightcurve versus ecliptic longitude of the comet-Sun vector, $\lambda_\sun$, under the assumption that the events are tied to specific illumination conditions on the nucleus.  Figure~\ref{fig:2003} compares the ZTF photometry to the 2003 MPC and NEAT data using absolute magnitude:
  \begin{equation}
    H(1,1,0) = m - 5 \log_{10}(\rh \Delta) - 2.5 \log_{10}(\Phi(\theta)),
  \end{equation}
where $\Phi$ is the Halley-Marcus phase function from \citet{schleicher11} evaluated at phase angle $\theta$.  Furthermore, the ZTF data are scaled to the 2003 circumstances using Eq.~\ref{eq:afrho}.  The pre-perihelion event would have spanned $\lambda_\sun=165-205$\degr{}, and its presence in 2003 cannot be tested.  The post-perihelion events observed in 2011 and 2018 would have spanned from $\lambda_\sun=305$\degr{} to at least 350\degr{}. Had this brightening occurred in the 2003 orbit, then we should have seen a comet near 18--19~mag, rather than the observed 20--21~mag.  Either the event did not occur at that time, or was much smaller in strength.

The 2003 lightcurve roughly agrees with $\afrho[0\degr]\sim40$~cm for $k=-6$ (Fig.~\ref{fig:2003}).  A model following $\afrho[0\degr]\sim$38~cm for $k=-4$ (not shown) fits equally well.  The lower activity level of this orbit is caused by the change in perihelion distance between 2003 and 2010.  Scaling the 2010 estimate, $\afrho[0\degr]\sim150$~cm, by $(2.53/2.12)^{-6}$ yields 52~cm at perihelion.  This factor of 3 change is larger than the factor of 2 predicted by the water ice sublimation model of \citet{cowan79}.

\section{Discussion}
The comet's behavior appears to have changed after the 2007 gravitational perturbation by Jupiter.  In 2010, the comet returned to perihelion with a peak activity level nearly consistent with the reduced perihelion distance.  180 days after perihelion, the comet brightened by $\sim-2$~mag, and remained bright for at least 90~days.  In 2017, the comet again brightened by $\sim-2$~mag, but this time slowly returned to a quiescent state, with a timescale much longer than is typical for outbursts, 150 versus 20--30 days \citep{hughes90}.  A third $-2$-mag brightening was observed in 2018, occurring in at least two stages over a 50-day period.  The comet remained at this new activity level for 190~days, up to the end of our data set.  Given these observations, we identify the following features of interest:

\begin{enumerate}
\item The comet brightened three times over two orbits, achieving nearly the same peak activity level each time: $\afrho[0\degr]\sim1350$~cm, corrected for heliocentric distance.  This behavior is unusual for cometary outbursts at short-period comets, which have a power-law distribution in total mass \citep{ishiguro16-outbursts}.  To have three large events of the same order of magnitude is an indication that the same active area may be responsible for all events.  A better understanding of this repeatability may provide insight into possible outburst trigger mechanisms, or the near-surface structure of the active area in question.

\item The three brightening events all occurred after the 2007 orbital perturbation by Jupiter, which increased surface insolation at perihelion by 40\% and the dust production rate by at least a factor of 3.  The data from the 2003 orbit are sparse, but there is no evidence for anomalous behavior.  Perhaps moderate changes in orbits can have profound consequences on cometary mass-loss.

\item Two of the brightening events occurred near the same point in the orbit on two separate orbits, near $T_P+180$~days in 2011 and 2018.  Both events are long lived, and the comet remains at the higher activity level for at least several months.  However, there is a lack of a similar event in 2003.  A single active area may be responsible for all events, and appears to have been in relative quiescence in 2003.
\end{enumerate}

We propose a scenario to account for these observations. The pre-2007 comet was near a steady state, balancing sublimation driven erosion with sub-surface devolatilization.  The perihelion distance change perturbed this scenario, and warmed volatile-rich sub-surface layers that were previously insulated from the thermal wave.  The new activity is isolated to one or two locations on the nucleus, indicating that the surface or immediate sub-surface is heterogeneous.

Terrain and activity heterogeneities are commonly observed by spacecraft missions to comets \citep{ahearn05, ahearn11, veverka13, thomas15-morph, hassig15}.  The first observed transition (2010 orbit, $T_P+182$~days) appears to have rejuvenated a local active area, perhaps shedding off an insulating layer.  We speculate that the same surface is illuminated upon the approach to perihelion.  Thus, the surface renewal may have occurred during the unobserved pre-perihelion approach in 2009.  Tests of this idea would benefit from a pole orientation measurement.  Thermophysical modeling of these events, and photometric observations during future perihelion passages will also help explore our proposed scenario.

Short-period comets, such as comet 240P, provide examples of cometary evolution from the cumulative effects of perihelion passages and orbital perturbations.  After perturbations to smaller perihelion distances, comets may have a greater tendency towards enhanced activity levels (cometary rejuvenation), or the volatile reservoirs may quickly diminish (rapid surface mantling or devolatilization).  Archival searches for comets under similar circumstances, and future comparisons to data taken with present-day surveys would benefit the study of comet behavior in general.  We observed comet 240P at an interesting moment in its evolution.  The pre-perihelion portion of the comet's 2025 orbit should be well-observed with the Large Synoptic Survey Telescope, with science operations expected to begin in 2023.

\acknowledgements

We thank all amateur astronomers contributing to the discovery of cometary outbursts.

Based on observations obtained with the Samuel Oschin Telescope 48-inch at the Palomar Observatory as part of the Zwicky Transient Facility project. ZTF is supported by the National Science Foundation under Grant No. AST-1440341 and a collaboration including Caltech, IPAC, the Weizmann Institute for Science, the Oskar Klein Center at Stockholm University, the University of Maryland, the University of Washington, Deutsches Elektronen-Synchrotron and Humboldt University, Los Alamos National Laboratories, the TANGO Consortium of Taiwan, the University of Wisconsin at Milwaukee, and Lawrence Berkeley National Laboratories. Operations are conducted by COO, IPAC, and UW.

\software{Astropy \citep{astropy18}, SEP \citep{barbary16-sep}, ZChecker \citep{Kelley2019}, Calviacat \citep{kelley19-calviacat}}
\facility{PO:1.2m (NEAT, PTF, ZTF)}
\bibliographystyle{aasjournal}
\bibliography{apj-jour,references,new-refs}

\begin{longrotatetable}
  \begin{deluxetable}{lcccccccccccc}
    \tablecaption{Photometry of comet 240P/NEAT.\label{tab:phot}}
  \tablehead{
    \colhead{Date}
    & \colhead{$T-T_P$}
    & \colhead{\rh} 
    & \colhead{$\Delta$}
    & \colhead{$\theta$}
    & \colhead{$N_{exp}$}
    & \colhead{$t_{exp}$}
    & \colhead{Airmass}
    & \colhead{Seeing}
    & \colhead{$\rho$}
    & \colhead{Filter}
    & \colhead{$m$}
    & \colhead{$\sigma_m$} \\
    \colhead{(UTC)}
    & \colhead{(days)}
    & \colhead{(au)} 
    & \colhead{(au)}
    & \colhead{(\degr)}
    & 
    & \colhead{(s)}
    & 
    & \colhead{(\arcsec)}
    & \colhead{(\arcsec)}
    & 
    & \colhead{(mag)}
    & \colhead{(mag)}
  }
\colnumbers
\startdata
\multicolumn{11}{l}{Near-Earth Asteroid Tracking Survey} \\\hline
2003-01-16 02:40 & -72.50 & 2.573 & 2.254 & 22.27 & 3 & 180 & 1.22 & 3.8 & 7.0 & $r$ & 18.52 & 0.07 \\
\hline
\multicolumn{11}{l}{Palomar Transient Factory} \\\hline
2010-07-25 11:42 & -70.78 & 2.196 & 2.575 & 22.85 & 1 & 60 & 2.30 & 2.7 & 9.50 & $r$ & 16.79 & 0.09 \\
2010-09-24 09:44 & -9.87 & 2.125 & 1.895 & 28.16 & 2 & 120 & 1.77 & 2.7 & 9.50 & $r$ & 16.05 & 0.06 \\
2010-12-05 11:11 & 62.19 & 2.180 & 1.271 & 13.21 & 1 & 60 & 1.06 & 2.1 & 9.50 & $r$ & 15.73 & 0.06 \\
2010-12-07 05:20 & 63.95 & 2.183 & 1.266 & 12.49 & 2 & 120 & 1.65 & 2.8 & 9.50 & $r$ & 15.69 & 0.04 \\
2010-12-08 05:17 & 64.95 & 2.184 & 1.263 & 12.09 & 2 & 120 & 1.64 & 2.5 & 9.50 & $r$ & 15.68 & 0.03 \\
\hline
\multicolumn{11}{l}{Zwicky Transient Facility} \\\hline
2017-12-12 02:46 & -154.76 & 2.442 & 2.235 & 23.77 & 1 & 180 & 1.90 & 3.7 & 9.26 & $r$ & 16.01 & 0.02 \\
2017-12-17 02:49 & -149.76 & 2.425 & 2.275 & 23.90 & 1 & 240 & 1.90 & 3.5 & 9.09 & $r$ & 16.06 & 0.02 \\
2018-09-11 12:22 & 118.64 & 2.324 & 2.935 & 17.62 & 1 & 30 & 1.98 & 2.5 & 7.05 & $r$ & 17.07 & 0.04 \\
2018-09-14 12:20 & 121.64 & 2.333 & 2.918 & 18.04 & 1 & 30 & 1.88 & 2.6 & 7.09 & $r$ & 17.14 & 0.04 \\
2018-09-16 12:09 & 123.63 & 2.339 & 2.907 & 18.30 & 1 & 30 & 2.31 & 2.3 & 7.12 & $g$ & 17.63 & 0.05 \\
2018-09-26 11:39 & 133.61 & 2.370 & 2.847 & 19.54 & 1 & 30 & 2.17 & 2.3 & 7.27 & $r$ & 17.25 & 0.07 \\
2018-09-29 11:49 & 136.61 & 2.380 & 2.828 & 19.89 & 1 & 30 & 1.90 & 2.8 & 7.32 & $g$ & 17.82 & 0.09 \\
2018-11-03 12:35 & 171.65 & 2.503 & 2.571 & 22.50 & 1 & 30 & 1.29 & 3.6 & 8.05 & $g$ & 17.40 & 0.05 \\
2018-11-03 11:37 & 171.61 & 2.503 & 2.572 & 22.50 & 1 & 30 & 1.59 & 3.7 & 8.05 & $r$ & 16.83 & 0.05 \\
2018-11-07 11:24 & 175.60 & 2.518 & 2.539 & 22.60 & 1 & 30 & 1.57 & 2.1 & 8.15 & $g$ & 17.45 & 0.04 \\
2018-11-07 12:34 & 175.65 & 2.519 & 2.539 & 22.60 & 1 & 30 & 1.24 & 1.7 & 8.15 & $r$ & 16.90 & 0.04 \\
2018-11-08 13:07 & 176.67 & 2.522 & 2.531 & 22.62 & 1 & 30 & 1.09 & 2.0 & 8.18 & $g$ & 17.47 & 0.04 \\
2018-11-10 11:48 & 178.61 & 2.530 & 2.515 & 22.64 & 1 & 30 & 1.37 & 2.4 & 8.23 & $g$ & 17.44 & 0.04 \\
2018-11-10 12:54 & 178.66 & 2.530 & 2.515 & 22.64 & 1 & 30 & 1.15 & 2.3 & 8.23 & $r$ & 16.90 & 0.03 \\
2018-11-16 11:10 & 184.59 & 2.553 & 2.466 & 22.65 & 1 & 30 & 1.44 & 2.0 & 8.39 & $g$ & 16.25 & 0.03 \\
2018-11-17 11:40 & 185.61 & 2.557 & 2.458 & 22.64 & 1 & 30 & 1.31 & 1.9 & 8.42 & $g$ & 16.19 & 0.03 \\
2018-11-19 11:58 & 187.62 & 2.565 & 2.441 & 22.60 & 1 & 30 & 1.20 & 3.0 & 8.48 & $g$ & 16.15 & 0.04 \\
2018-11-19 12:34 & 187.65 & 2.565 & 2.441 & 22.60 & 1 & 30 & 1.11 & 2.2 & 8.48 & $r$ & 15.59 & 0.03 \\
2018-11-25 12:10 & 193.63 & 2.588 & 2.392 & 22.42 & 2 & 60 & 1.16 & 2.3 & 8.65 & $r$ & 15.68 & 0.02 \\
2018-11-27 10:53 & 195.57 & 2.596 & 2.376 & 22.34 & 1 & 30 & 1.34 & 1.6 & 8.71 & $r$ & 15.76 & 0.04 \\
2018-11-28 12:07 & 196.63 & 2.600 & 2.367 & 22.29 & 1 & 30 & 1.13 & 2.3 & 8.74 & $g$ & 16.31 & 0.04 \\
2018-12-09 11:55 & 207.62 & 2.644 & 2.280 & 21.48 & 1 & 30 & 1.07 & 2.0 & 9.07 & $g$ & 16.44 & 0.03 \\
2018-12-09 12:52 & 207.66 & 2.645 & 2.280 & 21.48 & 1 & 30 & 1.01 & 2.1 & 9.07 & $r$ & 15.86 & 0.03 \\
2018-12-15 11:51 & 213.62 & 2.669 & 2.235 & 20.85 & 1 & 30 & 1.09 & 1.7 & 9.26 & $g$ & 16.40 & 0.07 \\
2018-12-15 12:54 & 213.66 & 2.669 & 2.235 & 20.84 & 2 & 60 & 1.01 & 2.4 & 9.26 & $r$ & 15.88 & 0.06 \\
2018-12-16 12:24 & 214.64 & 2.673 & 2.228 & 20.72 & 1 & 30 & 1.03 & 2.0 & 9.29 & $r$ & 15.87 & 0.02 \\
2018-12-22 11:29 & 220.60 & 2.698 & 2.186 & 19.93 & 1 & 30 & 1.08 & 1.9 & 9.46 & $r$ & 15.81 & 0.03 \\
2018-12-24 10:45 & 222.57 & 2.706 & 2.173 & 19.64 & 1 & 30 & 1.13 & 1.9 & 9.52 & $r$ & 15.76 & 0.03 \\
2018-12-28 11:55 & 226.62 & 2.723 & 2.148 & 18.99 & 1 & 30 & 1.02 & 2.5 & 9.63 & $g$ & 16.33 & 0.04 \\
2018-12-29 11:06 & 227.58 & 2.727 & 2.142 & 18.83 & 1 & 30 & 1.06 & 3.0 & 9.66 & $r$ & 15.75 & 0.03 \\
2018-12-30 13:00 & 228.66 & 2.731 & 2.136 & 18.64 & 1 & 30 & 1.00 & 2.1 & 9.69 & $g$ & 16.41 & 0.04 \\
2019-01-09 10:52 & 238.57 & 2.773 & 2.085 & 16.79 & 1 & 30 & 1.03 & 1.6 & 9.93 & $r$ & 15.89 & 0.04 \\
2019-01-10 10:42 & 239.57 & 2.778 & 2.080 & 16.59 & 1 & 30 & 1.04 & 1.8 & 9.95 & $r$ & 15.94 & 0.03 \\
2019-01-24 12:11 & 253.63 & 2.838 & 2.039 & 13.73 & 1 & 30 & 1.03 & 3.1 & 10.15 & $g$ & 16.52 & 0.05 \\
2019-02-08 10:57 & 268.58 & 2.902 & 2.040 & 11.29 & 2 & 60 & 1.05 & 2.5 & 10.14 & $g$ & 16.59 & 0.02 \\
2019-03-15 08:26 & 303.47 & 3.055 & 2.249 & 12.76 & 1 & 30 & 1.07 & 3.0 & 9.20 & $i$ & 16.29 & 0.04 \\
2019-03-25 08:27 & 313.47 & 3.099 & 2.358 & 14.19 & 1 & 30 & 1.14 & 1.9 & 8.77 & $i$ & 16.70 & 0.04 \\
2019-04-09 07:27 & 328.43 & 3.164 & 2.554 & 16.05 & 1 & 30 & 1.13 & 1.4 & 8.10 & $i$ & 17.16 & 0.05 \\
2019-04-12 05:51 & 331.37 & 3.177 & 2.596 & 16.34 & 1 & 30 & 1.02 & 1.6 & 7.97 & $i$ & 17.02 & 0.04 \\
2019-04-23 08:14 & 342.46 & 3.226 & 2.763 & 17.19 & 1 & 30 & 1.48 & 1.9 & 7.49 & $i$ & 17.45 & 0.06 \\
2019-05-09 04:00 & 358.29 & 3.295 & 3.019 & 17.71 & 1 & 30 & 1.02 & 1.4 & 6.85 & $i$ & 17.78 & 0.06 \\
2019-05-25 04:07 & 374.29 & 3.364 & 3.288 & 17.46 & 4 &120 & 1.08 & 1.9 & 6.29 & $i$ & 17.96 & 0.04 \\
\enddata
\tablecomments{Column definitions: (1) Mean observation date; (2) Time from nearest perihelion date; (3) heliocentric distance; (4) observer-comet distance; (5) phase angle (Sun-observer-target); (6) number of exposures; (7) total exposure time; (8) mean airmass; (9) mean seeing (stellar FWHM); (10) photometric aperture radius; (11) Filter bandpass, data from ZTF $r$ and PTF $R$ and NEAT (unfiltered) are calibrated to PS1 $r$, data from ZTF $g$ are calibrated to PS1 $g$; (12) apparent magnitude; (13) uncertainty on $m$.}
\end{deluxetable}
\end{longrotatetable}

\end{CJK*}
\end{document}